\documentstyle[twoside,amssymb]{article}
\newcommand{\R}{{\Bbb R}}
\newcommand{\C}{{\Bbb C}}

\newcommand{\bH}{{\bf H}}

\newcommand{\cB}{{\cal B}}

\newcommand{\cH}{{\cal H}}
\newcommand{\cI}{{\cal I}}

\newcommand{\cN}{{\cal N}}

\newcommand{\cV}{{\cal V}}
\newcommand{\ri}{{\rm i}}

\newcommand{\sQ}{{\sf Q}}

\newcommand{\Inf}{\mathop{\rm inf}}
\newcommand{\Sup}{\mathop{\rm sup}}

\newcommand{\Lim}{\mathop{{\rm lim}}\limits}

\newcommand{\dist}{\mathop{\rm dist}}
\newcommand{\Img}{\mathop{\rm Im}}
\newcommand{\Real}{\mathop{\rm Re}}

\newtheorem{theorem}{\sc Theorem}
\newtheorem{lemma}{\sc Lemma}

\title
{{OPERATOR INTERPRETATION OF RESONANCES ARISING IN SPECTRAL
      PROBLEMS FOR ${2}\times{2}$ MATRIX HAMILTONIANS%
\footnote{LANL E-print {\tt math-ph/9809016}. 
Contribution to Proceedings of the 
Conference ``Mathematical Results in Quantum Mechanics 
(QMath7)'', Prague, June 22-26, 1998. Published in Operator Theory: 
Advances and Applications {\bf 108} (1999), 315--322.}\,\,%
\footnote{Financial support of this work by the DFG, INTAS 
and RFBR is kindly acknowledged.}}}
\author
{A. K. Motovilov\thanks{Physikalishes Institut, Universit\"at Bonn,
      Endenicher Allee 11-13, D-53115 Bonn, Germany} \thanks{On leave
      of absence from the Joint Institute for Nuclear Research,
      141980 Dubna, Russia}\,,
R. Mennicken\thanks{Naturwiss. Fakult\"at I -- Mathematik,
      Universit\"at Regensburg,  D-93040 Regensburg, Germany}
}

\date{}
\begin{document}
\maketitle
\thispagestyle{empty}

\begin{quote}
We consider the analytic continuation of the transfer function
for a $2\times2$ matrix Hamiltonian into the unphysical sheets
of the energy Riemann surface. We construct non-selfadjoint
operators representing operator roots of the transfer function
which reproduce certain parts of its spectrum including resonances
situated in the unphysical sheets neighboring the physical
sheet. On this basis, completeness and basis properties for the
root vectors of the transfer function (including those for the
resonances) are proved.
\end{quote}
\bigskip

\noindent{\bf 1. Introduction.}
In this work we deal with Hamiltonian of the $2\times2$ matrix
form
$$
{\bf H}=\left(\begin{array}{lr}
  A_0        &   B_{01}  \\
  B_{10}     &   A_{1}
\end{array}\right)
$$
It is assumed that the operator $\bf H$ acts in an orthogonal
sum  ${\cal H}={\cal H}_0\oplus{\cal H}_1$ of the
Hilbert spaces ${\cal H}_0$ and ${\cal H}_1$ while the entries
$A_0:{\cal H}_0\rightarrow{\cal H}_0$, and $A_1:{\cal
H}_1\rightarrow{\cal H}_1$, are selfadjoint operators. The
couplings $B_{ij}:{\cal H}_j\rightarrow{\cal H}_i$, $i{\neq}j$,
$B_{01}=B^*_{10}$, are assumed to be bounded operators. We are
especially interested in the physically typical case where the
spectrum of, say, $A_1$ is partly or totally embedded into the
absolutely continuous spectrum of $A_0$ and the transfer
function $M_1(z)=A_1-z+V_1(z)$, where
$V_1(z)=B_{10}(z-A_0)^{-1}B_{01}$, admits analytic continuation
(as an operator-valued function) through the absolutely
continuous spectrum of the entry $A_0$ into the unphysical
sheet(s) of the energy $z$ plane. Notice that the resolvent
$({\bf H}-z)^{-1}$ of the operator $\bf H$ can be expressed
explicitly in terms of the inverse transfer function
$M_1^{-1}(z)$.  Therefore, in studying the spectral properties
of the transfer function $M_1$ one studies at the same time the
spectral properties of the operator matrix ${\bf H}$.

We construct an operator-valued function $V(X)$
(see~\cite{MotRem})
on the space of operators in ${\cal H}_1$ possessing
the property:  $V_1(X)\psi_1=V_1(z)\psi_1$ for any eigenvector
$\psi_1$ of $X$ corresponding to an eigenvalue $z$ and then
study the equation
\begin{equation}
\label{MainIni}
   H_1=A_1+V_1(H_1).
\end{equation}
In the case where the spectra of $A_0$ and $A_1$ have no
intersection, Eq.~(\ref{MainIni}) is reduced (cf.~\cite{MotRem})
to an operator Riccati equation for a similarity transform which
allows to find invariant subspaces of the matrix ${\bf H}$
admitting a graph representation~\cite{AdL,AdLMSr,MenShk} (also
see~\cite{MalyshevMinlos}).  Now we prove the solvability of
Eq.~(\ref{MainIni}) in the case where the spectra of $A_1$ and
$A_0$ overlap, cf.  also~\cite{MenShk}. However in this case the
solutions of~(\ref{MainIni}) already represent non-selfadjoint
operators.

The problem considered is closely related to the resonances
generated by the matrix $\bH$.  Regarding a definition of the
resonance and history of the subject see, e.\,g., the
book~\cite{ReedSimonIII}. A recent survey of the literature on
resonances can be found in~\cite{MotMathNach}. Throughout the
paper we treat resonances as the discrete spectrum of the
transfer function $M_1$ situated in the unphysical sheets of its
Riemann surface since points of this spectrum automatically
correspond to the poles of the analytic continuation of the
resolvent $({\bf H}-z)^{-1}$ (understood in the form sense).
Using the fact that the root vectors of the solutions $H_1$ of
the equation~(\ref{MainIni}) are at the same time such vectors
for $M_1$, we prove completeness and even basis properties for
the root vectors including those for the resonances. The results
obtained allow immediate applications in particular to the
scattering problems for multichannel Schr\"odinger equation.

A detailed exposition of the material presented including
proofs in the case of essentially more general
spectral situation will be given in the extended
paper~\cite{MennMot}.

\medskip

\noindent{\bf 2. Analytic continuation of the transfer function.}
For the sake of simplicity we assume in the report that all the
spectrum $\sigma(A_0)$ of the entry $A_0$ is absolutely
continuous consisting of an only interval
$J^0=(\mu_0^{(1)},\mu_0^{(2)})$ with
$-\infty\leq\mu_0^{(1)}<\mu_0^{(2)}\leq\infty$ while all the
spectrum $\sigma(A_1)$ of the entry $A_1$ is totally embedded
into the interval $J^0$.  Therefore, we assume
$\sigma(A_1)\subset J^0$.

Let $E_0$ be the spectral measure for the entry $A_0$,
$A_0=\int_{\sigma(A_0)}\lambda\,dE_0(\lambda)$.  Then the
function $V_1(z)$ can be written
$$
    V_1(z)=\displaystyle\int_{\sigma(A_0)}dK_B(\mu)(z-\mu)^{-1}
$$
with
$$
  K_B(\mu)=B_{10}E^0(\mu)B_{01}
$$
where $E^0(\mu)$ is the spectral function of $A_0$,
$E^0(\mu)=E_0\biggl((-\infty,\mu)\biggr)$. We suppose that
the function $K_B(\mu)$ is differentiable in
$\mu\in{J}^0$ in the operator norm topology. Further, we
suppose that the derivative $K'_B(\mu)$ is continuous within the
closed interval $\overline{{J}^0}$ and, moreover, that it
admits analytic continuation from this interval to a simply
connected domain situated, say, in $\C^-$. Let this domain be
called $D^-$. We assume that the boundary of $D^-$ includes the entire
spectral interval ${J}^0$.  Since $K'_B(\mu)$ represents a
selfadjoint operator for $\mu\in{J}^0$ and
${J}^0\subset\R$, the function $K'_B(\mu)$ also
automatically admits analytic continuation from ${J}^0$ into
the domain $D^+$, symmetric to $D^-$ with respect to the real
axis and $[K'_B(\mu)]^*=K'_B(\bar{\mu}),$
$\mu\in D^\pm\,.$

Let $\Gamma^l$, $l=\pm1$, be a rectifiable Jordan curve in
$D^l$ resulting from continuous deformation of the
interval ${J}^0$, the (finite) end points of this interval being
fixed. The quantity
$$\cV_0(B,\Gamma^l)=\displaystyle\int_{\Gamma^l}|d\mu|\,\|K'_B(\mu)\|$$
where $|d\mu|$ stands for the Lebesgue measure on $\Gamma^l$, is
called variation of the function $K_B(\mu)$ along the contour
$\Gamma^l$.  We suppose that there exists a contour (contours)
$\Gamma^l$ where the value $\cV_0(B,\Gamma^l)$ is finite
including also the case of the
unbounded interval ${J}^0$.  The contours $\Gamma^l$
satisfying the condition~$\cV_0(B,\Gamma^l)<\infty$ are said to
be $K_B$-bounded contours.

\begin{lemma}\label{M1-Continuation}
The analytic continuation of the transfer function $M_1(z)$,
$z\in\C\setminus{\sigma(A_0)}$, through the spectral interval
${J}^0$ into the subdomain $D(\Gamma^l)\subset D^l$,
$l=\pm1$, bounded by this interval and a $K_B$-bounded contour
$\Gamma^l$ is given by
\begin{equation}
\label{Mcmpl}
    M_1(z,\Gamma^l)=A_1-z+V_1(z,\Gamma^l)
    \quad\mbox{\rm with}\quad
V_1(z,\Gamma^l)=
\int_{\Gamma^l}
d\mu\,K'_B(\mu)\,(z-\mu)^{-1}.
\end{equation}
For $z\in D^l\cap D(\Gamma^l)$ one has
\mbox{$M_1(z,\Gamma^l)=M_1(z)+2\pi\ri\,l K'_B(z).$}
\end{lemma}
Proof is reduced to the observation
that the function $M_1(z,\Gamma^l)$ is holomorphic for
$z\in\C\setminus\Gamma^l$ and coincides with $M_1(z)$ for
$z\in\C\setminus\overline{D(\Gamma^l)}$.  The last equation
representing $M_1(z,\Gamma^l)$ via $M_1(z)$  is obtained
from~(\ref{Mcmpl}) using the Residue Theorem.

The latest statement of the lemma shows that the
transfer function $M_1$ has at least two-sheeted Riemann
surface.  The sheet of the complex plane $\C$ where the transfer
function $M_1(z)$ is considered together with the resolvent
\mbox{$({\bf H}-z)^{-1}$} initially is said to be the physical sheet.
The remaining sheets of the Riemann surface of $M_1$ are said
to be unphysical sheets. In the present work we only deal with
the unphysical sheets connected through the interval ${J}^0$
immediately to the physical sheet.
\medskip

\noindent{\bf 3. Example.}
To explain our reasons to introduce the function $K'_B(\mu)$ we 
will briefly consider an example closely related to the 
multichannel Schr\"odinger operator.  Let $\cH_0=L_2({\Bbb 
R}^n)$, $n\geq1$, and $A_0$ be the Laplacian, $A_0=-\Delta$, 
defined on the Sobolev space $W_2^2({\Bbb R}^n)$. We impose no 
restrictions on the Hilbert space $\cH_1$ and on the selfadjoint 
entry $A_1$ except for the assumption that 
$\sigma(A_1)\Subset\sigma(A_0)=[0,+\infty)$.  Let 
$\widetilde{b}\in L_2({\Bbb R}^n,\cH_1)$ be an $\cH_1$-valued 
function of the variable $x\in{\Bbb R}^n$ whose Fourier 
transform
$$
b(p)=
\displaystyle\frac{1}{(2\pi)^{n/2}}\int_{{\Bbb R}^n}
dx\,\,\widetilde{b}(x)\,\exp({\rm i}\langle p,x\rangle),
\quad p\in{\Bbb R}^n,
$$
represents a continuous function of $p$ with respect to the norm
topology in $\cH_1$. We define the coupling $B_{10}$ as
$$
B_{10}u_0=\displaystyle\int_{{\Bbb
R}^n}dx\,\widetilde{b}(x)u_0(x), \quad u_0\in\cH_0.
$$
Obviously, the operator $B_{10}$ is bounded and
$\|B_{10}\|\leq\|\widetilde{b}\|_{L_2({\Bbb R}^n,\cH_1)}$.  The
adjoint operator $B_{01}$ reads
$$
\bigl(B_{01}u_1\bigr)(x)=\langle
u_1,\widetilde{b}(x)\rangle_{\cH_1}, \quad u_1\in\cH_1.
$$

Denote by $\cB_r$ the open ball in ${\Bbb R}^n$ centered at the
origin and having radius $r$, $\cB_r=\{p\in{\Bbb
R}^n:\,|p|<r\}$.  For $\mu>0$ the value of the spectral
function $E^0(\mu)$ of the Laplacian $-\Delta$ represents
the integral operator in $L_2({\Bbb R}^n)$ whose kernel reads as
the Fourier transform of the characteristic function of the ball
$\cB_{\sqrt{\mu}}$ (see, e.\,g., Ref.~\cite{BirmanSolomiak},
\S4.2 of Ch.\,8),
$$
E^0(\mu;x,x')=\displaystyle\frac{1}{(2\pi)^n}\int_{\cB_{\sqrt{\mu}}}
dp\,\exp{({\rm i}\langle p,x-x'\rangle)}.
$$
If $\mu\leq0$ then $E^0(\mu;x,x')=0$.

A simple computation in the example considered shows that, for
$\mu>0$,
$$
   K'_B(\mu)=\displaystyle\frac{d}{d\mu}
   \biggl(B_{10}E^0(\mu)B_{01}\biggr)=
   \displaystyle\frac{1}{2}\,\mu^{(n-2)/2}
  \int_{S^{n-1}}d\widehat{p}\,\,\,b(\mu^{1/2}\widehat{p})\,\,
   \langle\,\cdot\,,b(\mu^{1/2}\widehat{p})\rangle_{\cH_1}
$$
where $S^{n-1}$ stands for the unit sphere in ${\Bbb R}^n$.  Thus,
for the analytic continuability of $K'_B(\mu)$ into a domain
$D\subset{\Bbb C}$ surrounding the spectrum of the entry $A_1$
it suffices to require the analytic continuability of the
function $b(p)$ into an appropriate domain of ${\Bbb C}^n$.  In
particular, if $b(p)$ admits an analytic continuation into a
``strip'' $|\Img p\,|<a$ for some $a>0$, then the function
$K'_B(\mu)$ is holomorphic in $\mu$ in the parabolic domain
$\Real\mu>-a^2+\displaystyle\frac{1}{4a^2}(\Img\mu)^2$ cut along
the interval $(-a^2,0]$. Notice that the holomorphy of $b(p)$ in 
a strip corresponds to the case of an exponentially decreasing  
$\widetilde{b}(x)$ as $x\to\infty$ which is often encountered in 
applications.

\medskip

\noindent{\bf 4. Factorization theorem.}
Let the spectrum of a linear operator $Y:\,\cH_1\to\cH_1$ is
separated from a $K_B$-bounded contour $\Gamma$\,. Then one can
define the operator
$$
V_1(Y,\Gamma)=\displaystyle
\int_{\Gamma}d\mu\,K'_B(\mu)(Y-\mu)^{-1}.
$$
Obviously, this operator is bounded,
$$
\|V_1(Y,\Gamma)\|\leq \cV_0(B,\Gamma)\,
\Sup_{\mu\in\Gamma}\|(Y-\mu)^{-1}\|\,.
$$
In what follows we consider the equation (cf.~\cite{MotRem})
\begin{equation}
\label{MainEqC}
X=V_1(A_1+X,\Gamma).
\end{equation}
This equation possesses the following important characteristic
property:  If $X$ is a solution of~(\ref{MainEqC}) and $u_1$ is
an eigenvector of \mbox{$H_1=A_1+X$}, $H_1 u_1=zu_1$,
then $zu_1=A_1 u_1+V_1(H_1,\Gamma)u_1=A_1
u_1+V_1(z,\Gamma)u_1.$ This implies that any eigenvalue $z$ of
$H_1$ is automatically an eigenvalue for the continued transfer
function $M_1(z,\Gamma^l)$ and $u_1$ is its eigenvector.  Thus,
having found the solution(s) of the equation~(\ref{MainEqC}) one
obtains an effective means of studying the spectral properties
of the transfer function $M_1(z,\Gamma)$ itself, referring
to well known facts of Operator Theory~\cite{GK,Kato}.
\begin{theorem}\label{Solvability}
Let a $K_B$-bounded contour $\Gamma$ satisfy the condition
\begin{equation}
\label{Best}
\cV_0(B,\Gamma)< \displaystyle\frac{1}{4}\,d_0^2(\Gamma)\,
\end{equation}
where $d_0(\Gamma)=\dist\{\sigma(A_1),\Gamma\}.$ Then
Eq.~{\rm(\ref{MainEqC})} is uniquely solvable in any ball
including operators $X:\cH_1\rightarrow\cH_1$ the norms of which
are bounded as $\|X\|\leq r$ with $r_{\rm min}(\Gamma)\leq
r<r_{\rm max}(\Gamma)$ where
\begin{equation}
\label{rminmax}
r_{\rm min}(\Gamma)=d_0(\Gamma)/2- \sqrt{d_0^2(\Gamma)/4
-\cV_0(B,\Gamma)}, \quad
r_{\rm max}(\Gamma)=d_0(\Gamma)-\sqrt{\cV_0(B,\Gamma)}. \,\,
\end{equation}
In fact, the solution $X$ belongs to the smallest ball
$\|X\|\leq r_{\rm min}(\Gamma)$.
\end{theorem}
One can prove this statement making use of the Banach's
Fixed Point Theorem (see~\cite{MennMot}).  One can
even prove that if the index $l=\pm1$ is fixed then,
under the condition~(\ref{Best}), the solution $X$
does not depend on
a concrete contour $\Gamma\subset D^l$.
Moreover, this solution satisfies the inequality
$\|X\|\leq r_0(B)$ with
\hbox{$
  r_0(B)=\Inf\limits_{\Gamma^l:\,\omega(B,\Gamma^l)>0}
  r_{\rm min}(\Gamma^l)\,
$}
where $\omega(B,\Gamma^l)=d_0^2(\Gamma^l)-4\cV_0(B,\Gamma^l).$
The value of $r_0(B)$ does not depend on $l$.  But when $l$
changes, the solution $X$ can also change.  For this reason we
supply it in the following with the index $l$ writing $X^{(l)}$.
As a matter of fact, the operators $H_1^{(l)}=A_1+X^{(l)}$,
$l=\pm1$, represent operator roots of the transfer function
$M_1$.
\begin{theorem}\label{factorization}
Let $\Gamma^l$ be a contour satisfying the
condition~{\rm(\ref{Best})} and $H_1^{(l)}=A_1+X^{(l)}$ where
$X^{(l)}$ is the above solution of the basic
equation~{\rm(\ref{MainEqC})}. Then, for
$z\in\C\setminus\Gamma^l$, the transfer function
$M_1(z,\Gamma^l)$ admits the factorization
\begin{equation}
\label{Mfactor}
    M_1(z,\Gamma^l)=W_1(z,\Gamma^l)\,(H_1^{(l)}-z)\,
\end{equation}
where $W_1(z,\Gamma^l)$
is a bounded operator in $\cH_1$,
$$
W_1(z,\Gamma^l)=I_1-\int_{\Gamma^l}
 d\mu\,K'_B(\mu)(H_1^{(l)}-\mu)^{-1}(\mu-z)^{-1}\,.
$$
Here, $I_1$ denotes the identity operator in ${\cal H}_1$.  For
$\dist\{z,\sigma(A_1)\}\leq{d_0(\Gamma^l)/2}$ the operator
$W_1(z,\Gamma^l)$ is boundedly invertible.

\end{theorem}
\begin{theorem}\label{SpHalfVic}
The spectrum $\sigma(H_1^{(l)})$ of the operator
$H_1^{(l)}=A_1+X^{(l)}$ belongs to the closed $r_0(B)$-vicinity
$$
{\cal O}_{r_0}(A_1)=
\{z\in\C:\,\,\dist\{z,\sigma(A_1)\}\leq
r_0(B)\}
$$
of the spectrum of $A_1$.  Moreover, the spectrum of
$M_1(\,\cdot\,,\Gamma^l)$ in
$$
{\cal O}_{d_0/2}(A_1)=\{z:\, z\in\C,\,
\dist\{z,\sigma(A_1)\}\leq{d_0(\Gamma^l)}/{2}\}
$$
is only represented by the spectrum of $H_1^{(l)}$ including the
complex spectrum {\rm(}in particular the resonances{\rm)}.
\end{theorem}

Let
$$
\Omega^{(l)}=\displaystyle\int_{\Gamma^l} d\mu\,
(H_1^{(-l)*}-\mu)^{-1}K'_B(\mu)\, (H_1^{(l)}-\mu)^{-1}
$$
where $\Gamma^l$ stands for a contour satisfying the
condition~(\ref{Best}).
\begin{theorem}
\label{MHOmega}
The operators $\Omega^{(l)}$, $l=\pm1$, possess the following
properties {\rm(}cf. {\rm\cite{MenShk}):}
\begin{eqnarray}
\nonumber
\|\Omega^{(l)}\|<1, \qquad
\Omega^{(-l)}&=&\Omega^{(l)*}, \\
\label{OmegaM}
-\frac{1}{2\pi\ri}\int_\gamma dz\,[M_1(z,\Gamma^l)]^{-1} &=&
(I_1+\Omega^{(l)})^{-1}\,,\\
\label{Hadj}
-\frac{1}{2\pi\ri}\int_\gamma dz\,z\,[M_1(z,\Gamma^l)]^{-1} &=&
(I_1+\Omega^{(l)})^{-1}H_1^{(-l)*}=
H_1^{(l)}(I_1+\Omega^{(l)})^{-1}\,\quad
\end{eqnarray}
where $\gamma$ stands for an arbitrary rectifiable closed
contour going in the positive direction around the
spectrum of $H_1^{(l)}$ inside the set ${\cal
O}_{d_0(\Gamma)/2}(A_1)$.  The integration over $\gamma$ is
understood in the strong sense.
\end{theorem}

The formulas~(\ref{OmegaM}) and~(\ref{Hadj}) allow one, in
principle, to find the operators $H_1^{(l)}$ and, thus, to
resolve the equation~(\ref{MainEqC}) only using the contour
integration of the inverse transfer function
$[M_1(z,\Gamma^l)]^{-1}$.  Also, Eq.~(\ref{Hadj}) implies that
the spectrum of $H_1^{(-)*}$ coincides with the spectrum of
$H_1^{(+)}$.

\medskip

\noindent{\bf 5. Completeness and basis properties.}
In the following we restrict ourselves to the case where the
resolvent $R_1(z)=(A_1-z)^{-1}$ of the entry $A_1$ is a compact
operator in $\cH_1$ for $z\in\rho(A_1)$. Then, according to
Theorem~IV.3.17 of~\cite{Kato}, the operators $H_1^{(l)}$ also
have compact resolvents since $X^{(l)}$ are bounded operators.

Denote by $\cH_{1,\lambda}^{(l)}$ the algebraic eigenspace of
$H_1^{(l)}$ corresponding to an eigenvalue $\lambda$.  Let
$m_\lambda$ be the algebraic multiplicity,
$m_\lambda=\mathop{\rm dim}\cH_{1,\lambda}^{(l)}$,
$m_\lambda<\infty$, and $\psi^{(l)}_{\lambda,i},$
$i=1,2,\ldots,m_\lambda,$ be the root vectors of $H_1^{(l)}$
forming a basis of the subspace $\cH_{1,\lambda}^{(l)}$.
Regarding these vectors we already have the following assertion
representing a particular case of Theorem~V.10.1 from~\cite{GK}.
\begin{theorem}\label{Completeness}
The closure of the linear span of the system
$\{\psi^{(l)}_{\lambda,i},\,\,\lambda\in\sigma(H_1^{(l)}),\,\,
i=1,2,\ldots,m_\lambda\}$ coincides with $\cH_1$.
\end{theorem}

We shall consider the case where the spectrum $\sigma(A_1)$
includes infinitely many points and the entry $A_1$ is
semibounded from below. The eigenvalues $\lambda_i^{(A_1)}$,
$i=1,2,\ldots\,$, of the operator $A_1$ are assumed to be
enumerated in increasing order;
\mbox{$\Lim_{i\to\infty}\lambda_i^{(A_1)}=+\infty$}.  Since we
assume $\sigma(A_1)\subset{J}^0$ this assumption means that
the interval ${J}^0$ is infinite, $\mu_0^{(2)}=\infty$.
Also, we suppose that there is a number $i_0$ such that for any
$i\geq i_0$
\begin{equation}
\label{DifLambda}
  \lambda_i^{(A_1)}-\lambda_{i-1}^{(A_1)}>2r.
\end{equation}
with some $r>r_0(B)$. Let $\gamma_0$ be a circle centered
at {$z=(\lambda_1^{(A_1)}+\lambda_{i_0-1}^{(A_1)})/2$} and
having the radius
{$(\lambda_{i_0-1}^{(A_1)}-\lambda_1^{(A_1)})/2+r$} while
$\gamma_i$ for $i\geq i_0$ be the circles centered at
$\lambda_i^{(A_1)}$ and having the radius $r$.  Let us introduce
the projections
$$
\sQ_i^{(l)}=-\frac{1}{2\pi\ri}\int_{\gamma_i}
dz\, (H_1^{(l)}-z)^{-1}, \quad i=0,i_0,i_0+1,\ldots\,.
$$
The subspaces \mbox{$\cN_i^{(l)}=\sQ_i^{(l)}\cH_1$} are 
invariant under $H_1^{(l)}$; $\mathop{\rm dim}\cN_i^{(l)}$ 
coincides with a sum of algebraic multiplicities of the 
eigenvalues $\lambda\in\sigma(H_1^{(l)})$ lying inside 
$\gamma_i$.
\begin{lemma}\label{NomegaLInd}
Under the condition~{\rm(\ref{DifLambda})} the sequence
$\cN_i^{(l)}$, \, $i=0,i_0,i_0+1,\ldots,$ \,
is $\omega$-linearly independent
and complete in $\cH_1$.
\end{lemma}
The next theorem represents a slightly extended statement of
Theorems~V.4.15 and~V.4.16 of~\cite{Kato} (the extension only
concerns a possible degeneracy of the eigenvalues of $A_1$).
\begin{theorem}\label{KatoV4-15-16}
Assume $\lambda_{i+1}^{(A_1)}-\lambda_i^{(A_1)}\to\infty$
as $i\to\infty.$ Let $i_0$ be a number starting from which
the inequality $\lambda_i^{(A_1)}-\lambda_{i-1}^{(A_1)}>4r_0$
holds.
Then the following limit exists
\begin{equation}
\label{sLimQ}
\mathop{s-\rm lim}\limits_{n\to\infty}
\sum_{i=0,i\geq i_0}^n \sQ_i^{(l)}=I_1.
\end{equation}
Additionally, assume that $\sum_{i=1}^\infty
{(\lambda_{i+1}^{(A_1)}-\lambda_i^{(A_1)})^{-2}}<\infty\,.$
Then~{\rm(\ref{sLimQ})} is true for any renumbering of
$\sQ_i^{(l)}$ and there exists a constant $C$ such that
$\biggl\|\sum_{i\in\cI}\sQ_i^{(l)}\biggr\|\leq C$ for any finite
set $\cI$ of integers $i=0$, $i\geq i_0$.
\end{theorem}
\bigskip

\baselineskip=12pt


\bigskip

\noindent{\small AMS Classification Numbers: Primary 47A56, 47Nxx;
                                           Secondary 47N50, 47A40.}
\end{document}